\documentclass{article}
\usepackage[T1]{fontenc} % add special characters (e.g., umlaute)
\usepackage[utf8]{inputenc} % set utf-8 as default input encoding
\usepackage{amsmath,cite,url}
\usepackage{graphicx}
\usepackage{color}
\usepackage{multirow}
\usepackage{float} 

\usepackage{tabularx}
\usepackage{multirow}
\usepackage{tabularx, multirow, array}

\newcolumntype{Y}{>{\centering\arraybackslash}X}
\title{musicolors: Bridging Sound and Visuals For Synesthetic Creative Musical Experience}

\author{
ChungHa Lee \\ 
Gwangju Institute of Science and Technology \\
\texttt{chunghalee@gm.gist.ac.kr}
\and
Jin-Hyuk Hong\thanks{Corresponding author.} \\
Gwangju Institute of Science and Technology \\
\texttt{jh7.hong@gist.ac.kr}
}

\begin{document}

\hbadness=99999  % or any number >=10000
\vbadness=99999  % or any number >=10000
\hfuzz=20pt
\maketitle
\begin{abstract}
Music visualization is an important medium that enables synesthetic experiences and creative inspiration. However, previous research focused mainly on the technical and theoretical aspects, overlooking users' everyday interaction with music visualizations. This gap highlights the pressing need for research on how music visualization influences users in synesthetic creative experiences and where they are heading. Thus, we developed \textbf{musicolors}, a web-based music visualization library available in real-time. Additionally, we conducted a qualitative user study with composers, developers, and listeners to explore how they use \textbf{musicolors} to appreciate and get inspiration and craft the music-visual interaction. The results show that \textbf{musicolors} provides a rich value of music visualization to users through sketching for musical ideas, integrating visualizations with other systems or platforms, and synesthetic listening. Based on these findings, we also provide guidelines for future music visualizations to offer a more interactive and creative experience.
\end{abstract}

\section{Introduction}\label{sec:introduction}
While music visualizations have traditionally served a range of objectives—including analysis \cite{1}, education \cite{2}, and therapy \cite{3}—their role in music appreciation is becoming increasingly significant \cite{4, 16, 112, 115}. They are now emerging as a critical component of digital entertainment and artistic expression \cite{5, 100}. This shift towards enjoyment emphasizes the importance of creating engaging and aesthetically pleasing visual experiences to accompany music. \textbf{\textit{Synesthesia}}, a perceptual phenomenon where stimulation in one sensory or cognitive pathway leads to automatic and involuntary experiences in another \cite{6, 22}, plays a crucial role in enriching experiences of users when they encounter music visualizations \cite{10, 11}. This integration of sensory modalities expands the music-listening experience, making music visualizations more captivating and visually appealing for users \cite{113, 114, 115}. Despite the benefits of synesthetic experiences in music visualizations, more user research is needed, particularly regarding how users react to and perceive potential in these visualizations.

Combining music with dynamic visuals also offers users a unique sensory experience for creative inspiration, inviting them to perceive and interact with music in visually expressive ways \cite{117, 118}. \textbf{\textit{Musical Creativity}} has been discussed with its potential and possibilities in the emotional \cite{116, 127} and conceptual perspectives \cite{126}, highlighting the importance of creative inspiration in the multimodal music system. However, there is a lack of research on which aspects of music visualization inspire users to be creative and what actions they can take with it. In this point, receiving diverse opinions from each user group using music visualizations is vital, suggesting the potential of music visualization systems from the users' perspective. Consequently, more research is required on what a different group of users of music visualization systems wants regarding synesthetic music appreciation and creative inspiration. Based on this research gap, our \textbf{contributions} are as follows.

\begin{itemize}
\item The development of \textbf{musicolors}, a real-time music visualization library that is web-based and easy for users to use.
\item A qualitative user study that compares and provides insights into the experiences of different user groups when using the \textbf{musicolors}.
\item Guidelines for future music visualization regarding synesthetic experience and creative inspiration.
\end{itemize}

To understand the users' perspective on synesthetic creative music visualizations, we developed \textbf{musicolors}, a JavaScript library designed to visualize pitch, energy, and timbre in real-time. The qualitative user study with eight participants, including composers, developers of musical systems or APIs, and music listeners, showed that our system evoked a synesthetic experience and creative minds by playing the instrument or streaming music online with \textbf{musicolors}. We also offer guidelines to system developers, designers, and researchers for providing inventive inspiration and a synesthetic experience with the music visualization system.

\section{Related Work}\label{sec:related_work}

\subsection{Music Visualizations}\label{subsec:music_visualizations}
Extensive research combining music with visualization systems has focused on integrating different information and navigation schemes to represent elements of music \cite{9, 10}, including rhythm\cite{104}, melody \cite{105, 106}, and tempo \cite{4}. For instance, \textit{MoshViz} enhanced the understanding of music renditions by focusing on specific musical instruments and providing a detailed overview visualization approach\cite{13}. \textit{Visual Harmony} represented harmonic structures to assist composers in efficiently composing music pieces\cite{14}. While these studies have demonstrated the effectiveness of the proposed visualization systems through user studies, the purpose of the systems is focused on composing or music analysis, which is far from the purpose of appreciation. This gap shows that users' demand for recognizing instantaneous changes to music and ways to maximize the user's experience has also increased parallelly \cite{120, 121}.

Most real-time music visualization systems for appreciation have been made for performance. Bain focused on creating 3D animations as visual accompaniments to live music performances, enhancing the concert experience by translating music into visual narratives\cite{100}. Still, these visualizations require advanced hardware and software, making them less accessible for low-budget productions. Su and Ling proposed an immersive design for music visualization in performance spaces tailored to electronic music, utilizing mapping rules and technical means to visualize music\cite{18}. Their approach, however, was primarily suited for electronic music. Robyn and Torres introduced a music visualization system that uses responsive imagery to create interactive visual experiences based on live performance feature data, focused on usage with art performers\cite{19}. Thus, most systems were geared toward the perspective of the artist or performer on the stage or under specific circumstances, with advanced hardware or software. These studies also show the noticeable gap in music visualization technologies that are accessible and user-friendly for non-performers. Hence, there is a clear need to redirect efforts towards developing the system with the broader perspective of users in mind, aiming to make music visualization a more integral part of daily music experiences.

\subsection{Synesthesia and Creativity in MIR Systems}\label{subsec:synestheisa_creativity}
Recent discussions within the MIR community have focused on the efforts to link musical features to visual elements to facilitate synesthetic experiences \cite{113, 114, 115}. Still, the existing studies on music-visual synesthesia have primarily explored the theoretical aspect of blending sounds with visual experiences, such as mapping rules. For example, O'Neill proposed a theoretical foundation for mapping fundamental visual/music parameters such as pitch, energy, and timbre \cite{21}. Similarly, Mao et al. introduced a theoretical model of visualized timbre synesthesia (ISCM) and a music synesthesia visualization tool (ASHA) \cite{22}. These advancements have paved the way for innovative tools that enhance synesthetic creativity and expression across diverse artistic domains. Nevertheless, there is a noticeable gap in research on users' synesthesia in the context of visualizations, indicating a need for further exploration of the perception and potential of these visualizations. The scarcity of papers examining these experiences from the users' perspective underscores the need for more in-depth accounts of how individuals experience synesthesia through music visualizations.

Regarding creativity, Collins et al. developed a piano-roll web interface, suggesting that future research should assess whether such technology could enhance musical creativity\cite{23}. Andersen and Knees emphasized the importance of fostering enhanced flow and extended periods of concentration and creativity\cite{24} but primarily focused on music experts specializing in electronic dance music. These studies highlight the need for an interdisciplinary approach in designing MIR systems that creatively represent musical information, a point also highlighted by Humphrey et al.\cite{25}. While previous research has concurred that users can creatively leverage musical information and has advocated for the necessity of related research or systems, a gap exists in studies that have directly explored how various users utilize specific systems and the potentialities therein. Therefore, we sought to develop a music visualization system for users to observe how they engage with music synesthetically and how the system creatively inspires them.

\begin{figure*}[ht]
  \centering
  \includegraphics[width=\linewidth]{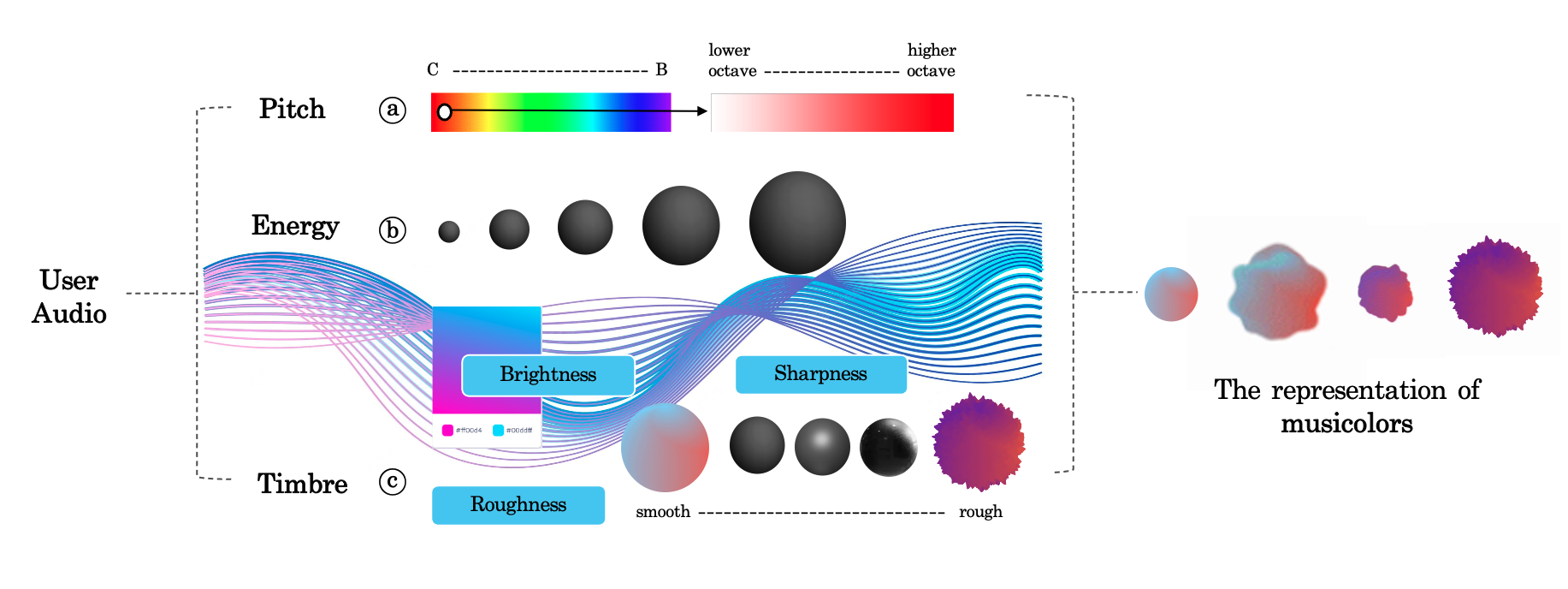}
  \caption{The Overall Framework of \textbf{musicolors}}
\end{figure*}

\section{Design}\label{sec:design}
This section describes the user groups and tasks, mapping rules with the framework of \textbf{musicolors}, and the interface for processing user audio. Figure 1 illustrates the overall framework of our system.

\subsection{User Groups and Tasks}\label{subsec:user_tasks}
Our goal was to ascertain if \textbf{musicolors} could enable different users to experience synesthesia or boost their creativity and discover valuable insights from each group. Therefore, we categorized participants into three user groups based on their identities: Composer, Music system/API Developer, and Listener. 

Due to the division between the three groups, we designed the experiment so each group could take on different tasks. Tasks for the Composer and Developer group were designed to use \textbf{musicolors} with their ongoing musical projects or to contemplate the connection between their work and the system. Conversely, the Listener group's tasks focused on integrating \textbf{musicolors} into their daily routines and naturally discovering points of interaction with music.

\subsection{The Mapping Rules}\label{subsec:mapping_rules}
We employed Three.js\footnote{https://threejs.org/}, a JavaScript library renowned for its three-dimensional visualization capabilities, to make our system intuitive. As a foundational visual element, we selected a sphere shape, which serves as the base for dynamic alterations synchronized with changes in the music. Our mapping scheme presents the three primary musical elements—\textit{Pitch, Energy, and Timbre}—and the corresponding visual elements we associated with each.

\subsubsection{Pitch and Colors}\label{subsubsec:pitch_colors}
The correlation between pitch and color is delineated through real-time analysis of the audio input's fundamental frequency in \textbf{musicolors}. Employing the Pitchy\footnote{https://www.npmjs.com/package/pitchy}, the system translates each detected pitch into its corresponding musical note and octave. Based on the previous studies about pitch and color mappings for synesthesia \cite{21, 26, 103}, the system assigns a distinct color to each note from a carefully curated palette for visual representation (Figure 1a). This color assignment scheme mirrors the note's position on the musical scale, facilitating an intuitive visual mapping. For example, colors representing lower octaves are chosen for their lower saturation, whereas higher octaves are depicted with higher saturation.

\subsubsection{Energy and Size}\label{subsubsec:energy_size}
Energy, represented by the audio signal's loudness, is quantified utilizing the \textit{energy} feature from the Meyda\footnote{https://meyda.js.org/}—a JavaScript audio feature extraction library \cite{111}. This metric dynamically measures the loudness of the audio, reflecting the sound's intensity at any moment. In the \textbf{musicolors}, energy's visual representation is directly linked to the size of a three-dimensional object, referenced by previous research about energy and size mappings \cite{4, 21}. The energy metric significantly influences the object's scaling factor, with heightened energy levels yielding larger visualization (Figure 1b). This mechanism of size modulation articulately conveys the audio energy visually, providing a tangible representation of the sound's intensity.

\subsubsection{Timbre and Textures}\label{subsubsec:timbre_textures}
Timbre is visually rendered through an integration of audio feature extractions from Meyda, including features of \textit{perceptualSpread}, \textit{spectralFlatness}, and \textit{spectralKurtosis}. These attributes encapsulate the sound's timbral complexity, translating it into visual textures and patterns. The \textbf{musicolors} employs timbral characteristics to modulate the texture of the rendered object (Figure 1c) based on the reference papers about timbral features and textural feelings \cite{21, 28, 29}. For instance, \textit{spectralFlatness} contributes to surface \textit{roughness}, engendering intricate textures for harmonically rich sounds. Concurrently, \textit{perceptualSharpness} influences the visual definition and \textit{sharpness}, amplifying the object's luminance to signify sharper sounds. Moreover, the \textit{spectralCentroid} feature, indicative of the sound's \textit{brightness}, modulates the object's hue, harmonizing the visual color spectrum with the auditory spectral content.

\subsubsection{The Details for Music-Visual Mappings Selection}\label{subsubsec:mapping_details}
In the design of \textbf{musicolors}, mapping musical elements to visual attributes was critical for ensuring an intuitive and engaging user experience. Our selection process involved considering various alternatives before settling on the most effective mappings. 

For pitch, while previous studies considered mapping to motion \cite{122, 123}, color was selected due to its ability to convey pitch variations instantly. This choice was supported by a prior study on synesthetic experiences with pitch and colors \cite{26} (\textit{e.g.}, where higher pitches are often associated with lighter colors). Previous attempts for energy mapping presented the option of thickness or stroke \cite{124}, but size was selected for its precise representation of loudness or intensity due to its intuitiveness to use \cite{21}. This direct correlation made \textbf{musicolors} represent music's dynamics in real-time, enhancing the visualization's immersive experience. Timbre has also been expressed through shapes \cite{125} or motion \cite{119}. However, the texture was the most fitting because it was connected directly with the semantic descriptors related to the texture feelings of music \cite{36}, offering a rich layer of detail that shapes or motion could oversimplify.

\subsection{User Audio Processing and Interface}\label{subsec:audio_process}
The \textbf{musicolors} library facilitates real-time visualization within a web browser, leveraging the \textit{getUserMedia}\footnote{https://developer.mozilla.org/en-US/search?q=getUserMedia} method for direct audio input access. This capability ensures that users can freely visualize sounds from any source captured by their device's microphone, including voices, musical instruments, and ambient sounds. It can also be readily integrated into web projects, offering developers a straightforward method to incorporate real-time audio visualization into their applications. 

On the user's side, \textbf{musicolors} autonomously processes and visualizes the audio input without necessitating external software or specialized hardware. Figure 2 shows the web-based user interface of the \textbf{musicolors} with the usage examples, such as streaming music (Figure 2a) or playing an instrument (Figure 2b). Our system's code implementation can also be found at [Anonymized link].

\begin{figure}[ht]
  \centering
  \includegraphics[width=0.7\linewidth]{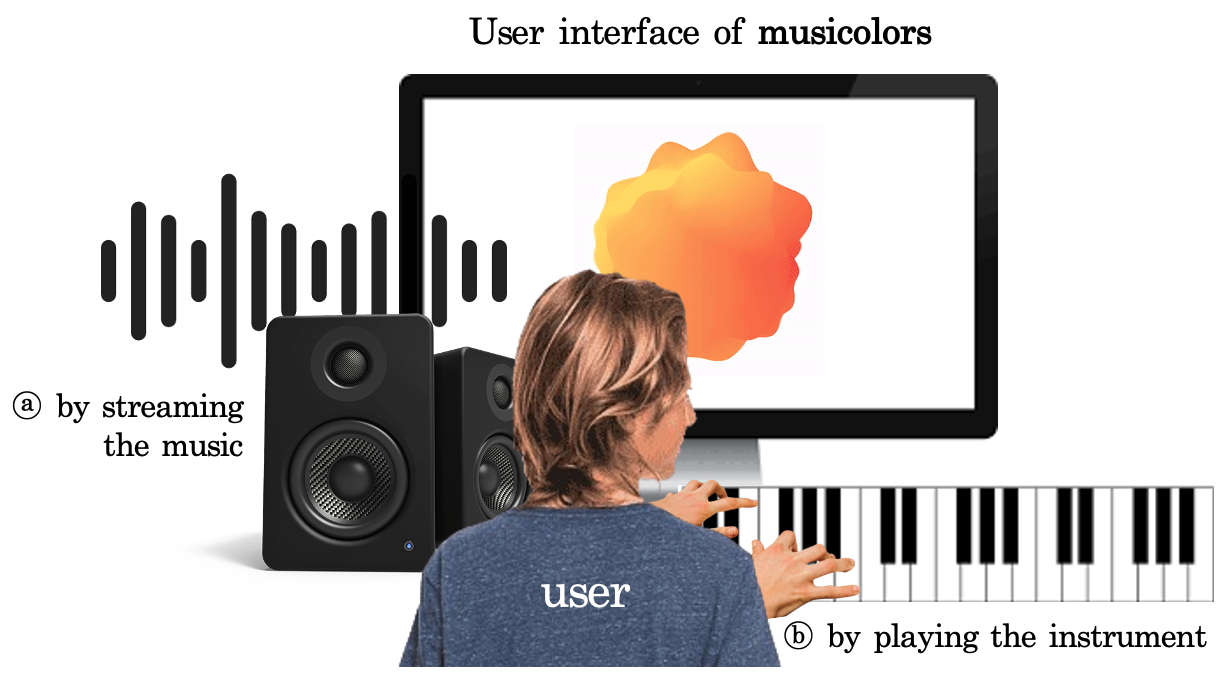}
  \caption{The User Interface and Examples of \textbf{musicolors}}
\end{figure}

\section{User Study}\label{sec:user_study}

\subsection{Study Process}
We chose a qualitative user study to explore individuals' in-depth stories using music visualizations and uncover their experiences and insights. After we recruited participants through snowball sampling, we were able to compose each group, as Table 1 shows. For the experiment, we asked each participant to use \textbf{musicolors} for two days, allowing them to engage with the tool freely based on their musical preferences and listening styles. 

Since the participants were divided into three groups, we tailored the interview questions to consider their occupation and musical knowledge. After using the system, we interviewed each participant about user experience, new insights compared to the previous music listening experience, and potential improvements for the system. After the experiment, each participant was compensated with 10,000 KRW in cash. Following the interviews, the data was transcribed for thematic analysis \cite{30}. The lead author initially coded these transcripts using an open inductive approach, followed by discussions with the research team for analysis.

\begin{table}[h]
 \centering
 \begin{tabularx}{\columnwidth}{Y|>{\centering\arraybackslash}p{1.25cm}|>{\centering\arraybackslash}p{1.25cm}|>{\centering\arraybackslash}p{2cm}}
  \hline
  \textbf{Group} & \textbf{ID} & \textbf{Age} & \textbf{Sex} \\
  \hline
  \hline
  \multirow{2}{*}{Composer} & P1 & 33 & Female \\
                            & P2 & 31 & Male \\ \hline
  \multirow{2}{*}{Developer} & P3 & 25 & Male \\
                             & P4 & 26 & Male \\ \hline
  \multirow{4}{*}{Listener} & P5 & 28 & Male \\
                            & P6 & 25 & Male \\
                            & P7 & 26 & Male \\
                            & P8 & 23 & Female \\ \hline
 \end{tabularx}
 \caption{Demographic Information of Participants}
 \label{tab:table1}
\end{table}

\subsection{Results}

\subsubsection{Bridging Sound and Visuals: Composers' Insights on Creative Minds} \label{subsec:results1} 

\textbf{The potential for the initial creative material on sound textures.} We let all users use the \textbf{musicolors} without explaining the mapping rules; all composers commented that the visualization matched the feel of the music they had in mind, especially the timbre part. P2, who had 13 years of experience composing music, thought that \textbf{musicolors} express timbre well through texture expression. \textit{``When the instruments were switched up in the music, I saw the textures also switching up by the instrument's timbre.''} Additionally, we found that composers view \textbf{musicolors} as an earlier extension of the materials and tools for composing music or the sound itself. P1, who had 14 years of specialty in composing music, recognized the potential of using \textbf{musicolors} in the composing process for amateurs. \textit{``Typically, composers rely on analyzers to examine sound textures, but I think \textbf{musicolors} offer an intuitive way for amateur composers or the general public to feel the sound.''} These results show that our system has the potential to provide creative inspiration based on the texture of sound or to create other creations through inspiration that connects sound and visuals.

\textbf{The medium for inspiration by musical kinesthetic and the context.} Both composers noted that users with expertise in music or composition would desire \textbf{musicolors} from two perspectives: (1) \textit{musical kinesthetics} and (2) \textit{musical context}. Regarding musical kinesthetics, P1 recounted her experience with a sound installation display, highlighting that \textbf{musicolors} could be a valuable kinesthetic aid in sound exhibitions. \textit{``The sound exhibitions are qualified as aural art. Incorporating visuals in this setting can enhance its accessibility, so I am sure \textbf{musicolors} will superbly capture music's dynamic essence in sound exhibitions.''} P2 mentioned that \textbf{musicolors} could aid in preserving the musical context through visuals, as these remain in the user's view by enabling verification of audio through visuals twice. \textit{``Since music comprises auditory information that fades over time, a visual representation of sound progression in \textbf{musicolors} can assist artists in retaining the music's context.''} Our results show that it is vital to keep the kinetics and context of music in the visualization system like \textbf{musicolors}, which will be critical points for the creative activities.

\vspace{2pt}
\subsubsection{Crafting Interaction: Developers' View on Making a Synergistic Music Visualization} \label{subsubsec:results2}

\textbf{The adaptability with other music API or multimodal systems.} We discovered that developers utilized \textbf{musicolors}, focusing on the overall system design, users' flow, and visual design of the system. Developers found the connection point of \textbf{musicolors} with the system they had developed beforehand. P3, who had experience developing APIs for real-time analysis and analytics of music elements, mentioned that \textbf{musicolors} could be integrated with the APIs he previously developed. P4, who developed a system providing genre-specific haptic filters, stated that combining visual and haptic representation would anticipate users with synesthesia when delivering music. \textit{``If the visualization and haptics are well synchronized, it will result in a synergistic effect, offering users a more comprehensive sensory experience.''}

\textbf{Enhancing the users' flow and access to experience musical dynamics.} Most importantly, developers anticipated that users of \textbf{musicolors} would experience novel sensations akin to metaphors or new symbols. They argued that users' experiencing music with visualizations is associated with \textit{`musical dynamics'}, which gives \textbf{musicolors} an advantage over other music visualization systems that work with MIDI or WAV files. This advantage comes from its ability to transfer information and enhance users' listening experience in real-time. Along with these, developers also suggested the further application of the system, putting together a Bluetooth speaker with an immersive LED display or an art object that reacts automatically to the sound or music as P3 stated: \textit{``Since it is a web-based library, it would be very accessible not just to developers but also to people who just want to attach \textbf{musicolors} anywhere and anytime they want.''}

\textbf{The aesthetic visual design to capture the attention.} Another aspect our developers found captivating about \textbf{musicolors} is its aesthetic appeal. Both developers were particularly drawn to the novelty of its aesthetics, which sets it apart from current music visualizations in the market. As P4 stated, \textit{``I think this system is unique because it delivers a comprehensive presentation into a cohesive display, with the 3D shape of a sphere integrating color, size, and texture.''} P3 also commented on using color gradients and geometric shapes that respond to the user audio: \textit{``These visuals can get attention from the users because they change dynamically by the sound. Users are naturally drawn to such visualizations, catching their eyes even as they pass by.''}

\vspace{2pt}
\subsubsection{Shaping Synesthesia: Enhancing Music listeners' Appreciation with \textbf{musicolors}} \label{subsubsec:results3}

\textbf{Touching the live imagination of listening experience.} While the four music listeners were less aware of the mapping relationships in \textbf{musicolors} than the composers and developers, they were eagerly curious about the musical elements represented, focusing on the intuitive changes in color, size, and texture. This curiosity eventually led users to focus on the sound's features with the visuals in the system. In particular, P6 and P8 noted that the music felt \textit{`more alive'} than when they usually listened to music in audio. P5 reflected on the experience of listening to music with \textbf{musicolors} as follows: \textit{``The part where the music visualization moved dynamically according to the rhythm or the change in the sound of the instruments was where I found it fun to use \textbf{musicolors}.''} Additionally, P7, who listened to the band's dynamic fusion of jazz and modern electronic music, mentioned that while he usually imagined the band playing when listening to music alone, experiencing the music through the \textbf{musicolors} made his music experience more vivid and colorful.

\textbf{Synesthetic experience with vividly expressed sound features.} As recalled by the participants, these results suggest a synesthetic experience, where the auditory and visual experiences are merged. While these results align with other studies proposing music visualization systems \cite{31, 32, 33, 34}, the unique visuals of \textbf{musicolors} enhanced the music listening experience by enabling listeners to immerse themselves fully. In particular, P6 stated, \textit{``The visualization expanding and contracting in sync with the repetitive drum sounds was quite eye-opening. I think I could see the drum sounds and the bass line of the music better when I listened to the music with \textbf{musicolors}.''} Similarly, P8 recalled that \textbf{musicolors} allowed them to focus more on the music because the visualization intuitively changed according to the intensity of the beats and sounds. Consequently, we found out that the visuals in our system evoked the vivid experience of synesthesia, making participants link their sound experience to the visuals.

\textbf{Suggestions of the new eras on music visualizations.}
In addition to using \textbf{musicolors} for enjoyment, music listeners also suggested ways to use it for sharing their personal music experiences with the music visualization. P5 said that if custom features were added to the system, he would express his feelings for the music he wants to recommend to others. \textit{``I think we could use these visualizations to express something melancholic that is difficult to express in words.''} P8 also suggested, \textit{``It could be linked to social media on streaming platforms like Spotify, where you can use accustomed \textbf{musicolors} to share and display your music appreciation with others.''} These results showed that users are looking for a visual medium to explain and share their appreciation with others, leading us to discuss the possibilities of the new culture of music visualizations.

\section{Discussion}\label{sec:discussion}
Our goal was to conduct a user study to understand how composers, music system/API developers, and listeners perceive and utilize \textbf{musicolors}, a real-time music visualization system. We recruited participants with diverse backgrounds to compare the perspectives of composers seeking creative inspiration, developers aiming to grasp the system architecture and user intentions, and users desiring a synesthetic music experience. We believe this study offers valuable insights into the varied uses and potential desires for music visualization systems. Given these insights, we propose the following guidelines for developers, designers, and researchers interested in creating systems akin to \textbf{musicolors}.

First, for the music visualization systems to inspire creativity effectively, they need to define and communicate the various elements of music intuitively alongside diverse visual representations. This implies that professionals, amateur artists, and users looking to draw inspiration from music for other creative ventures must identify common symbols or metaphors that are understandable to all. Efforts in this direction can build on research into semantic descriptors of music \cite{35, 36} or metaphorical definitions of timbre sensations \cite{37}. As research on semantically defining music elements progresses, techniques for more intuitive mappings between musical and visual elements will evolve.

Secondly, to transform music visualization into a synesthetic experience, developing systems and integrating them into music-related content or streaming platforms as user participants highlighted in the Results section is crucial. Adding audiovisual content to platforms where users commonly listen to music will not only enrich the music-listening experience but also provide a novel experience where the visualization alters the perception of the music, even when the same track is replayed.

Finally, it is essential to expand the concept of music visualization beyond passive listening into a medium for users to share their musical impressions. Our research indicates that users looked forward to expressing their feelings about music in challenging ways to convey through text alone. Therefore, incorporating user-based custom features into music visualization systems and enabling their sharing with others will foster a vibrant culture of \textit{`sharing music visualizations'}.

\section{Limitations and Future Work}\label{sec:limitation_futurework}
While our study offers valuable insights into the potential of \textbf{musicolors} for enhancing the music listening experience through real-time visualization, it is essential to acknowledge certain limitations warrant further exploration.

Firstly, the participant sample size in our user study was small, suggesting future research could involve a more extensive and varied group of participants to ensure the findings reflect a broader range of user experiences and cultural backgrounds.

Additionally, the current version of \textbf{musicolors} lacks customizable features, which can allow users to personalize their visual music experience. Our participants wanted such functions to share their unique musical experiences and interpretations with others. This highlights a significant opportunity for enhancing user engagement and interaction within the \textbf{musicolors}. Corporating user-customizable features will facilitate a more personal and shared experience, promoting a more profound connection among users over their musical tastes and preferences.

Lastly, the mapping between musical elements (such as pitch, energy, and timbre) and their visual representations (color, size, and texture) remains an area that could benefit from further research. The subjective nature of music and visual art makes establishing a universally accepted set of mapping rules challenging. Our study relied on intuitive mappings that may hold different meanings or significance for all users. Future studies should explore the audiological and psychological underpinnings of music-visual synesthesia, exploring how these mappings can be optimized to enhance the synesthetic creative experience for a broader audience.

By addressing these limitations, future work can build upon our initial findings to create more inclusive, customizable, and scientifically grounded music visualization experiences that cater to a broader audience and enrich the musical landscape.

\section{Conclusion}\label{sec:conclusion}
In this study, we developed \textbf{musicolors}, a system that visualizes music's pitch, energy, and timbre in real-time. We conducted a qualitative user study with eight participants, including composers, music system/API developers, and listeners. Our qualitative user study with composers showed that \textbf{musicolors} can serve as a medium for innovative inspiration for users with creative minds. A study with developers found insights on how to craft visual interaction with the system, focusing on the adaptability, users' flow, and aesthetics of \textbf{musicolors}. From the music listeners, we found that visual features in our system made a synesthetic experience with the new suggestions on music visualization. Our results demonstrated that \textbf{musicolors} provides users with a rich value of music visualizations, from sketching musical ideas to synesthetically enjoying the music. Moreover, we suggest that music visualization be actively integrated into music-related platforms and streaming sites, with the potential for a rich and colorful culture where users share their personal feelings through music visualization.

\bibliographystyle{plain} % 또는 unsrt, ieee 등
\bibliography{ref}

\end{document}